# Learning Apparent Diffusion Coefficient Maps from Accelerated Radial k-Space Diffusion-Weighted MRI in Mice using a Deep CNN-Transformer Model

Yuemeng Li[1,2], Miguel Romanello Joaquim[2], Stephen Pickup[2], Hee Kwon Song[2], Rong Zhou[2,3], Yong Fan[1,2]*

[1]Center for Biomedical Image Computing and Analytics (CBICA), Perelman School of Medicine, University of Pennsylvania, Philadelphia, PA 19104, USA
[2]Department of Radiology, Perelman School of Medicine, University of Pennsylvania, Philadelphia, PA 19104, USA
[3]Abramson Cancer Center, University of Pennsylvania, Philadelphia, PA 19104, USA

*Corresponding author:
Name:         Yong Fan, Ph.D.
Department:   Department of Radiology
Institute:    Perelman School of Medicine, University of Pennsylvania
E-mail:       Yong.Fan@pennmedicine.upenn.edu

Approximate word count: 194 (Abstract) 5000 (body)

## Abstract:

**Purpose:** To accelerate radially sampled diffusion weighted spin-echo (Rad-DW-SE) acquisition method for generating high quality apparent diffusion coefficient (ADC) maps.

**Methods:** A deep learning method was developed to generate accurate ADC maps from accelerated DWI data acquired with the Rad-DW-SE method. The deep learning method integrates convolutional neural networks (CNNs) with vision transformers to generate high quality ADC maps from accelerated DWI data, regularized by a monoexponential ADC model fitting term. A model was trained on DWI data of 147 mice and evaluated on DWI data of 36 mice, with acceleration factors of 4x and 8x compared to the original acquisition parameters. We have made our code publicly available at GitHub: https://github.com/ymli39/DeepADC-Net-Learning-Apparent-Diffusion-Coefficient-Maps, and our dataset can be downloaded at https://pennpancreaticcancerimagingresource.github.io/data.html.

**Results:** Ablation studies and experimental results have demonstrated that the proposed deep learning model generates higher quality ADC maps from accelerated DWI data than alternative deep learning methods under comparison when their performance is quantified in whole images as well as in regions of interest, including tumors, kidneys, and muscles.

**Conclusions:** The deep learning method with integrated CNNs and transformers provides an effective means to accurately compute ADC maps from accelerated DWI data acquired with the Rad-DW-SE method.

# 1   Introduction

Diffusion weighted (DW) MRI provides quantitative metrics related to the Brownian motion of water hindered by microstructures present in biological tissues (1,2). The apparent diffusion coefficient (ADC) of water derived from DW images at multiple *b*-values has been employed extensively as a biomarker in neurological and oncological applications (3-6). Accurate ADC map generation can assist in differentiating between benign and malignant tumors, determining tumor aggressiveness and monitoring tumor response to treatment (7,8). Furthermore, accurate ADC assessment with reducing scan time has several advantages, including better subject tolerance of shorter scans and potential reduction of motion-related artifacts which are exacerbated by long scan times. Since DWI pulse sequences are sensitive to motion on the micrometer scale, macroscopic (millimeter) scale movement of tissue /organ due to respiratory motion can introduce artifacts, resulting in errors in quantitative measurement of ADC in the affected tissue. In clinical DWI, respiratory motion is often mitigated by breath-holds or with respiratory navigators, as well as by employing rapid acquisition schemes such as single-shot EPI (9) to minimize motion corruption. In DWI of mice, however, higher respiration rates and increased magnetic susceptibility effects due to high magnetic field strength, EPI-based DWI performed on preclinical MRI instruments leads to greater levels of distortions and artifacts (10). By leveraging the intrinsic motion-insensitive property of radial k-space sampling, previous studies have shown that the radially sampled diffusion weighted spin-echo (Rad-DW-SE) acquisition method effectively suppresses respiratory motion artifacts in DW-MR images of mouse abdomen over a wide range of *b*-values (10,11). Nevertheless, compared to single-shot EPI, the acquisition time of Rad-DW-SE is substantially longer. An effective means to shorten the Rad-DW-SE scanning time is to acquire under-sampled k-space data. However, accelerating Rad-DW-SE k-space data acquisition degrades the image quality dramatically, especially at higher *b*-values due to lower signal-to-noise ratios (SNR), and subsequently degrades the derived ADC maps.

Two approaches can be adopted to generate high quality ADC maps from accelerated DW images: 1) generating high quality DW images followed by fitting to a diffusion model to estimate ADC (12); or 2) directly generating high quality ADC maps from accelerated ADC maps. High quality DW images can be generated using deep learning methods that have achieved promising performance from accelerated k-space data in k-space domain (13-15), image domain (16-21), or both (22-24). However, the performance of such indirect methods hinges on the quality of the generated DW images at different *b*-values with varied signal-to-noise ratios. On the other hand, directly generating high quality ADC maps from under-sampled ADC maps can be implemented using a deep learning model under a supervised learning setting. However, such an approach only utilizes ADC maps and does not utilize individual DW images.

In this study, we develop a deep learning model, referred to as DeepADC-Net, to generate high quality ADC maps from radially accelerated DW data, in conjunction with a monoexponential diffusion model that estimates the ADC maps from the DW images. Our deep learning model takes the accelerated DW images of multiple *b*-values and their derived ADC map as a multi-channel input to generate high quality ADC maps. The deep learning model is trained to optimize two complementary loss functions 1) the difference between the ADC maps generated by deep learning model and those derived from fully-sampled DW images using a monoexponential model, and 2) the difference between fully-sampled DW images and the DW images estimated from the learned ADC maps (12). Different from the existing deep learning based MR image generation methods that are typically built on convolutional neural networks (CNNs), our deep learning method is an integration of CNNs with vision transformers (25) that have shown great potential to learn the global context information as a self-attention module in conjunction with CNNs for feature extractions (26). Extensive ablation studies and experimental results demonstrate that the monoexponential model and the integration of CNNs with vision transformers could enhance deep learning to generate high quality ADC maps from accelerated data. Although many deep learning methods have been developed for MRI data generation tasks with different image acquisition methods, our method is developed to improve ADC computation from accelerated DW data collected with the radial k-space sampled diffusion weighted spin-echo (Rad-DW-SE) acquisition method.

## 2 Methods

### 2.1 Datasets

All animal handling protocols were reviewed and approved by the IACUC of the University of Pennsylvania. Animal studies employed a genetically engineered mouse model of pancreatic ductal adenocarcinoma that spontaneously develops premalignant pancreatic intraepithelial neoplastic lesions at 7–10 weeks of age. Animals were prepared for MRI exam by induction of general anesthesia and placement of vital signs probes as detailed elsewhere (10). MRI was performed on a 9.4T horizontal bore scanner (Bruker, Billerica, MA) using a 35 mm quadrature birdcage RF coil (M2M, Cleveland, Oh) for transmit and reception (10). Following a set of localizers, a contiguous series of axial slices spanning the tumor volume were acquired using a diffusion weighted, radially sampled spin echo sequence (Rad-DW-SE) with even sampling of the view angles over 360 degrees by acquiring one spoke per echo (FOV=$32 \times 32 mm^2$, slices= $8 - 19, 96$ readout points, $403$ views, slice thickness=$1.5\ mm$, TR=$750\ msec$, TE=$28.7\ msec$, $b$-values= $10, 535, 1070, 1479, 2141\ s/mm^2$, total acquisition time=$25\ min$). To achieve sufficient SNR and image quality, 403 spokes were used as the reference acquisition, although Nyquist criterion requires fewer views (~150). Subsequently, 4x and 8x reduction in sampling from the reference acquisition was used to evaluate the effectiveness of our deep learning strategy for accelerating the data acquisition.

Furthermore, we also collected "real-world" accelerated data at a factor of 4x and compared ADC maps generated by our deep learning model with those computed from fully-sampled data that were subsequentially collected using the same protocol. It is worth noting that the real-world 4x accelerated data might capture diffusion information different from that captured by the full-sampled data in that they were not collected simultaneously. All relevant results are presented in the supplemental document.

Based on the fully sampled Rad-DW-SE data, we evaluated our proposed model with two different acceleration factors of 4x and 8x. The 4x accelerated DW data were generated by sampling one out of every four radial views in k-space, resulting in a total of 101 views, while the 8x accelerated DW data were generated by sampling one out of every eight radial views in k-space, resulting in a total of 50 views. DW images were reconstructed from both the fully sampled and accelerated k-space data using Python code developed in house. Following zero and first order phase correction of each of the acquired views, the k-space data were regridded to a 96x96 Cartesian array using the Kaiser-Bessel kernel and a convolution window of four. The regridded data were then Fourier transformed and divided by the deconvolution function to yield the reconstructed images.

ADC maps were computed from both the accelerated and the fully sampled DW images by least-squares-fitting the monoexponential model (27). The ADC maps derived from the fully sampled DW images were used as ground truth, and values were excluded if they lie outside of the range [0, 0.0032] mm$^2$/sec since 0.0032 corresponds to ADC of free water at 37°C (28). We split the entire dataset into a training subset with scans of 147 animals consisting of a total of 2255 slices, and a testing subset with scans of 36 animals with a total of 557 slices.

## 2.2 DeepADC-Net

### 2.2.1 Problem formulation

Given Rad-DW-SE scans collected at $n$ different *b*-values, an ADC map can be computed from the DW images by fitting a monoexponential model $\mathcal{M}$:

$$S_i = \mathcal{M}(ADC, S_0) = S_0 e^{-b_i \times ADC}, \quad i = 1, \ldots, n, \quad (1)$$

where $S_i$ is intensity value of a DW scan at a b-value of $b_i, i = 1, \ldots, n$, $ADC$ is the ADC parametric map, and $S_0$ is the intensity value of a DW scan in the absence of diffusion weighting. Based on the monoexponential model, the ADC map can be calculated from DW images collected with at least two different *b*-values:

$$ADC_{fit} = -\frac{\ln(S_j) - \ln(S_i)}{b_j - b_i}, \quad (2)$$

where $b_i$ and $b_j$ are two different b-values, and $S_i$ and $S_j$ are their corresponding DW images. To make the ADC estimation robust, DW images are typically collected at three or more *b*-values, and a least-squares-fitting algorithm is then adopted to estimate the ADC and $S_0$ values.

Given accelerated DW images, we aim to optimize the DeepADC-Net to generate high-quality ADC maps close to those estimated from their corresponding full-sampled DW images:

$$\max_{\theta} Similarity(ADC_{full}, ADC_{cnn} = F_{CNN}(S_i, ADC_{us}|\theta)), \quad (3)$$

where $ADC_{cnn} = F_{CNN}(S_i, ADC_{us}|\theta)$ is a deep learning model with parameter $\theta$, its input consists of the accelerated DW images $S_i, i = 1, ..., n$ and their corresponding ADC map $ADC_{us}$ computed by fitting the monoexponential model, $ADC_{full}$ is the fully sampled ADC map computed using the monoexponential model from their corresponding fully sampled DW images, and $Similarity(\cdot,\cdot)$ is a similarity measure between two ADC maps.

### 2.2.2 DeepADC Network architecture

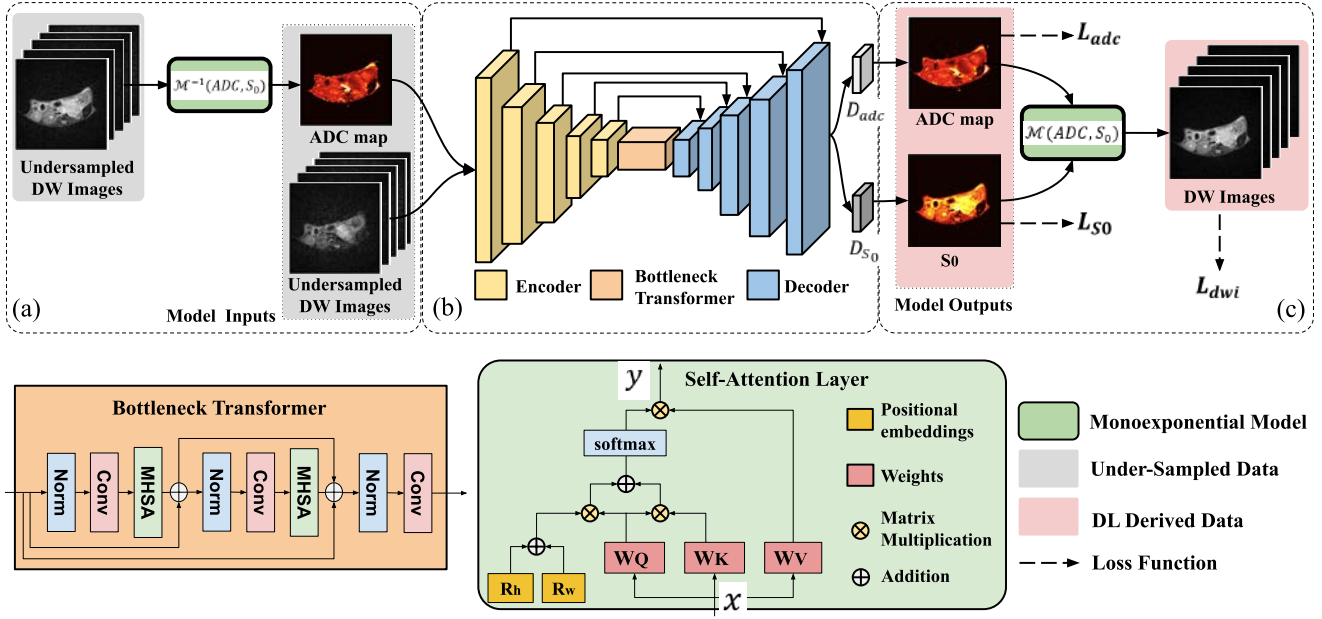

Figure 1. DeepADC-Net flowchart: a) the input consists of multiple channels, including the accelerated DW images and their corresponding ADC maps generated by fitting a monoexponential diffusion model; b) a densely connected encoder-decoder backbone that contains a bottleneck transformer with the self-attention; c) the output includes high quality ADC map and $S_0$ where high quality DW images are generated from those outputs using the monoexponential model.

DeepADC-Net is constructed to generate high quality ADC maps from accelerated DWI data collected with the Rad-DW-SE sequence, as schematically illustrated in Figure 1. The input to the deep learning model includes the accelerated DW images and their corresponding ADC maps. The model consists of two parts in the training setting: 1) using accelerated DW images and their corresponding ADC maps as multi-channel input, as illustrated in Figure 1(a), to generate high quality ADC maps and $S_0$ map, which corresponds to a DW image in the absence of diffusion weighting (12), as shown in Figure 1(b); 2) estimating high quality DW images from generated ADC and $S_0$ with the monoexponential model, illustrated in Figure 1(c). In the inference setting, the deep learning model is applied to accelerated DW images and their corresponding ADC maps to generate high quality ADC maps, as shown in Figure 1 (a) and (b).

**Encoder-Decoder architecture:** In deep learning tasks, the encoder compresses the input data into a low-dimensional representation and the decoder generates high-dimensional output data from the low-dimensional representation, while U-Net is a type of Encoder-Decoder network (29) that includes skip connections between the encoder and the decoder (30). In our approach, high quality ADC maps are generated using an Encoder-Decoder network with five densely connected blocks for both the encoder and the decoder (31). This network takes both the accelerated DW images $DWI_{us} \in \mathbb{R}^{H \times W \times n}$ at $n$ b-values and their corresponding ADC map $ADC_{us} \in \mathbb{R}^{H \times W \times 1}$ as a multi-channel input, where $H$ and $W$ are the dimensions of the image matrix. The Decoder's last layer has two paralleled output heads $D_{adc}$ and $D_{S_0}$ that generate a high quality ADC map $\widetilde{ADC} \in \mathbb{R}^{H \times W \times 1}$ and its corresponding DW image at a b-value of 0, denoted by $\widetilde{S_0} \in \mathbb{R}^{H \times W \times 1}$, respectively.

To generate ADC values within a physiologically plausible range, we adopt a scaled sigmoid activation function, in the decoder's output head $D_{adc}$, formulated as:

$$\widetilde{ADC} = ADC_{min} + \frac{1}{1+e^{-X_{adc}}} \times (ADC_{max} - ADC_{min}), \qquad (4)$$

where $ADC_{min}$ and $ADC_{max}$ are the lower and upper boundaries of ADC values respectively, and $X_{adc} \in \mathbb{R}^{H \times W \times 1}$ is the output of $D_{adc}$ (i.e., the one before the activation layer). $ADC_{min}$ was set to 0, which is the smallest possible value indicating no diffusion, and $ADC_{max}$ was set to the ADC of free water at 37°C, which equals 0.0032 $mm^2/s$ and is presumably the largest possible value in vivo.

According to the monoexponential model specified in Equation (1), both ADC and $S_0$ are learnable parametric maps, where $S_0 \in \mathbb{R}^{H \times W \times 1}$ represents the intensity values of the DW image in absence of diffusion weighting, and DW images $S = [S_1, ..., S_n] \in \mathbb{R}^{H \times W \times n}$ are collected with diffusion weighting at $n$ b-values. The intensity values of DW images at different b-values are positive, decreasing as the b-value increases. Therefore, the output of $D_{S_0}$ should be equal to or larger than its corresponding DW scan acquired at the lowest b-value, denoted by $S_1$. Accordingly, the output of $D_{S_0}$ is formulated as:

$$\widetilde{S_0} = (1 + Max(0, X_{S_0})) \times S_1, \qquad (5)$$

where $S_1$ is the DW scan collected at the lowest b-value, and $X_{S_0} \in \mathbb{R}^{H \times W \times 1}$ is a feature map generated by the decoder's output head $D_{S_0}$.

Given $\widetilde{ADC}$ and $\widetilde{S_0}$ generated from the deep learning model and their ground truths $ADC$ and $S_0$ generated from the monoexponential model using fully-sampled DW images, DW images at different b-values can be calculated to regularize the output ADC maps $\widetilde{ADC}$ by encouraging the generated DW images $\tilde{S} = [\widetilde{S_1}, ..., \widetilde{S_n}] \in \mathbb{R}^{H \times W \times n}$ to be close to their corresponding fully sampled DW images $S = [S_1, ..., S_n] \in \mathbb{R}^{H \times W \times n}$. According to Equation (1), DW images $\tilde{S}$ can be computed with the generated

$\widetilde{ADC} \in \mathbb{R}^{H \times W \times 1}$ and $\widetilde{S_0} \in \mathbb{R}^{H \times W \times 1}$ by the monoexponential model, referred to as $\mathcal{M}(ADC, S_0)$, at $n$ b-values used for collecting the DW images.

**Bottleneck Self-Attention:** Attention mechanism is a technique used in deep learning to selectively focus on different parts of the input data, allowing deep learning models to capture the relationships between different elements of the input and make more informed decisions (32). The bottleneck self-attention is a variant of self-attention that specifically focuses on the bottleneck of an Encoder-Decoder deep learning network (26). Let $x \in \mathbb{R}^{C_{in} \times H \times W}$ be the input feature map, $q = W_Q x$ be the queries, $k = W_k x$ be the keys, and $v = W_V x$ be the values, the output $y \in \mathbb{R}^{C_{out} \times H \times W}$ from self-attention layer can be computed as:

$$y_{ij} = \sum_{h=1}^{H} \sum_{w=1}^{W} \text{softmax}(q_{ij}^T k_{hw}) v_{hw}, \tag{6}$$

where $i \in \{1, ..., H\}$ and $j \in \{1, ..., W\}$ represent two different locations, $q_{ij}$, $k_{hw}$, and $v_{hw}$ are elements of learned attention weights $W_Q \in \mathbb{R}^{C_{in} \times C_{out}}$, $W_k \in \mathbb{R}^{C_{in} \times C_{out}}$, and $W_v \in \mathbb{R}^{C_{in} \times C_{out}}$, respectively. To make the self-attention mechanism sensitive to the positional information, the learnable position encoding is incorporated into the self-attention layer. The Multi-Head self-attention module is applied by taking different query, key, and value matrices to enable the attention layers focus on different parts of the input feature maps.

**Loss functions:** Multiple loss functions are adopted to optimize the network for generating high quality ADC maps, including:

$$L_{adc} = \frac{1}{N} \sum_{j=1}^{N} \| ADC_j - \widetilde{ADC}_j \|_1,$$

$$L_{S_0} = \frac{1}{N} \sum_{j=1}^{N} \| S_{0_j} - \widetilde{S_0}_j \|_1, \tag{7}$$

$$L_{dwi} = \frac{1}{N} \sum_{j=1}^{N} \| S_j - \widetilde{S}_j \|_1,$$

where $\widetilde{ADC}_j$ is the generated ADC map and $\widetilde{S_0}_j$ is DW images at b-value of 0, $\widetilde{S}_j$ are the DW images computed from generated $\widetilde{ADC}_j$ and $\widetilde{S_0}_j$ using $\mathcal{M}(\widetilde{ADC}_j, \widetilde{S_0}_j)$ with $n$ b-values according to equation (1), and $ADC_j$, $S_{0_j}$ and $S_j$ are their counterparts of the fully sampled data. The overall loss function is:

$$L_{total} = \alpha L_{adc} + \beta L_{S_0} + \gamma L_{dwi}, \tag{8}$$

where $\alpha, \beta,$ and $\gamma$ are regularization parameters. We set $\alpha = 1$ and $\beta = \gamma = 0.1$, which yielded the overall best results among a range of values, as detailed in Supplementary Table S2 and Section 1.2.

## 2.3 Implementation Details and Evaluation Metrics

We performed our experiment on a single NVIDIA TITAN RTX GPU with PyTorch implementation. We utilized the Adam optimizer with a learning rate of $1 \times 10^{-5}$, and a weight decay of $1 \times 10^{-4}$. We chose the head size of four for multi-head self-attention module (26) in our implemented bottleneck transformer. The model was trained in a total of 1000 epochs in approximately two hours. While the dynamic range of the signal intensities of DW images with five b-values are highly varied, its corresponding ADC maps are in the range of [0, 0.0032] $mm^2/s$, where the maximal value corresponds to ADC of free water at 37°C. To normalize the DW images and their corresponding ADC maps into feasible scales to feed into the deep learning model, we therefore clipped the DWI data with a maximum value of 99 percentile, and further normalized the clipped DWI data into a [0, 1] range. The ADC maps were normalized into the range of [0, 1] during training and the predicted ADC maps were scaled back to their original ranges.

We evaluated ADC map generation based on the testing images. Particularly, ADC maps computed from the fully sampled imaging data were used as ground-truth data. To quantitatively evaluate the generated ADC maps, we used correlation coefficient (CC) for quantifying linear relationship of voxelwise values between the generated and ground truth ADC maps, in addition to structural similarity (SSIM) index, peak signal-to-noise ratio (PSNR), and normalized mean square error (NMSE), which are widely adopted in image generation studies (22-24). We evaluated the generated ADC maps of the testing data, focusing on the whole images and regions of interests (ROIs), including tumor, muscle, and kidney. To evaluate the ADC on the whole image basis, we utilized all testing images with background excluded by masking out the non-tissue regions to reduce the influence of noise during the evaluation. Instead of manually generating the tissue masks, we automatically excluded image pixels if their ADC values were outside of [0, 0.0032] or their intensity values of the DWI scans were not in descending order according to their ascending b-values. The ROIs were manually labeled.

### 2.4 Comparison with state-of-the-art deep learning and compressed sensing methods

We compared our DeepADC-Net with state-of-the-art deep learning methods, including: 1) U-Net (30), 2) DenseU-Net (31), 3) FBP-ConvNet (16), and 4) Att-UNet (17). These methods were implemented with the same network architectures as reported in their corresponding papers to generate high quality ADC maps from the accelerated ADC maps. We utilized the same training and inference setting to train all the models, where the best models were saved based on the best correlation coefficient score estimated based on the training dataset, and we evaluated the models' overall performance on the testing dataset, with all four quantitative metrics, including CC, SSIM, PSNR and NMSE as described in Section 2.3. Specifically, the U-Net model contained encoders and decoders, each with 4 convolutional blocks; the DenseU-Net model contained encoders and decoders, each with 5 densely-connected blocks; the FBP-ConvNet model used the U-Net based architecture with a skip connection between the input and the output; and the Att-UNet model utilized a channel attention mechanism within the U-Net backbone. Furthermore, we also compared our method with a compressed sensing method (33) by adopting an implementation provided by SparseMRI V0.2[1]. The CS method was implemented using MATLAB (version R2022a, MathWorks, Natick, MA). Since the CS method performs best when using a randomized sampling scheme, we selected the views for the accelerated datasets using block randomization, where a random view is selected out of every four.

## 2.5 Ablation studies

Table 1. Ablation studies of deep learning models trained with (✓) and without (✗) indicated components of the proposed deep learning method, including different inputs, different loss functions, and self-attention.

| Models | inputs | $L_{adc}$ | $L_{S0}$ | $L_{dwi}$ | Self-Attention |
|---|---|---|---|---|---|
| DenseU-Net | ADC | ✓ | ✗ | ✗ | ✗ |
| DenseU-ADC | ADC+DWI | ✓ | ✗ | ✗ | ✗ |
| DenseU-DWI | ADC+DWI | ✗ | ✗ | ✓ | ✗ |
| DenseU-ADC-DWI | ADC+DWI | ✓ | ✓ | ✓ | ✗ |
| DeepADC-Net | ADC+DWI | ✓ | ✓ | ✓ | ✓ |

By systematically removing individual components of DL processing to determine their contribution to the overall performance, we carried out ablation studies to investigate how different components of the proposed deep learning method contribute to the ADC map generation, including different inputs, different combinations of the loss function term, and self-attention, as summarized in Table 1. All the models with different components under evaluation had the same DenseU-Net backbone (31), one model had the input of ADC map generated from the accelerated DW images by fitting the monoexponential model, and all the other models had the same multi-channel input of both the accelerated DW images and their associated ADC map. All the models were trained and evaluated with the same training and inference settings. All the studies were performed on the accelerated DW images with an acceleration factor of 4.

### 2.5.1 Ablation studies on network inputs

As indicated in Table 1 (rows 1 and 2), we evaluated how different inputs contributed to the ADC generation with the deep learning models trained by optimizing $L_{adc}$. Specifically, the DenseU-Net model with the accelerated ADC map alone as its input is referred to as DenseU-Net, whereas the DenseU-Net model with the multi-channel input of both the accelerated DW images at multiple b-values and their ADC map is referred to as DenseU-ADC.

### 2.5.2 Ablation study on loss functions

---
[1] http://people.eecs.berkeley.edu/~mlustig/Software.html

We evaluated how the loss function terms contributed to the ADC generation with the deep learning models trained with the multi-channel input of both the accelerated DW images at multiple b-values and their ADC map. Specifically, we trained two deep learning models to generate ADC maps and $S_0$, which are used to further compute DW images by optimizing $L_{dwi}$ or a combination of $L_{adc}$, $L_{S_0}$ and $L_{dwi}$. The former is referred to as DenseU-DWI and the latter is referred to as DenseU-ADC-DWI.

### 2.5.3 Ablation study on self-attentions

We also investigated how the self-attention module contributed to the generation of ADC maps. The multi-head self-attention (MHSA) module was implemented in the network bottleneck, consisting of three densely connected convolutional layers, and the MHSA module was applied after the first and second convolutional layers. The model is referred to as DeepADC-Net and was compared with DenseU-ADC-DWI.

## 3 Results

### 3.1 Comparison with state-of-the-art Methods on the Entire Cross-Section

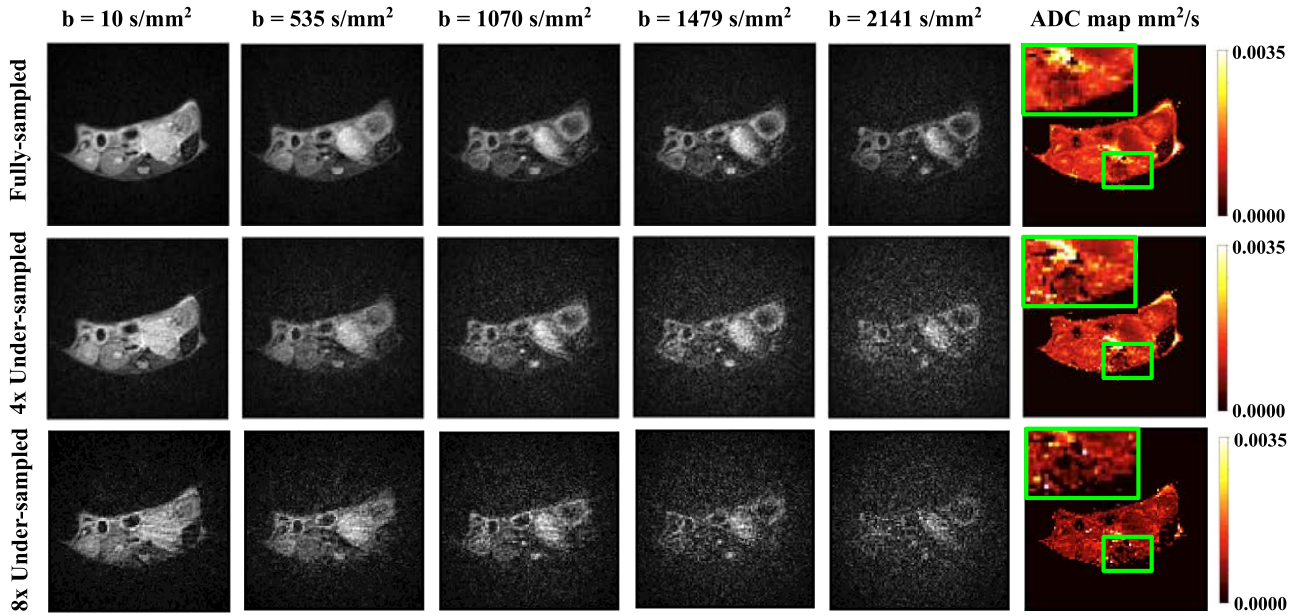

Figure 2. Diffusion weighted images and ADC maps from fully and accelerated Rad-DW-SE scans at different *b*-values. The accelerated images were obtained by down-sampling the fully-sampled data with acceleration factor of four and eight, and the ADC maps were computed from their corresponding DW images by fitting a monoexponential model. Compared with their fully-sampled counterparts, the accelerated DW images appeared noisy, especially at higher b-values, and the derived ADC maps lost anatomical detail as shown in green bounding boxes. The ADC decreases as the degree of acceleration increases.

Figure 2 shows representative fully-sampled and accelerated DW images and their corresponding ADC maps obtained by fitting the monoexponential model, indicating that the accelerated DW images were noisy, especially at higher b-values, and the derived ADC maps lose anatomical details.

Table 2 summarizes quantitative evaluation metrics for ADC maps generated by DeepADC-Net and alternative state-of-the-art methods under comparison. The ADC maps estimated from accelerated DW images with least-squares-fitting were substantially different from those derived from their corresponding fully sampled DW images. U-net, DenseU-Net, Att-UNet, and FBP-ConvNet yielded ADC maps with improved similarity to fully sampled data compared with those estimated directly from accelerated DW images with least-squares-fitting. DeepADC-Net yielded the best similarity for all metrics studied. DeepADC-Net improved upon the second-best method by 3.15%, 0.13%, 1.29dB, and 0.63% in terms of CC, SSIM, PSNR, and NMSE, respectively. The quantitative evaluation results summarized in Table 2 also demonstrate that it was more challenging to estimate high quality ADC maps from the 8x accelerated data than from the 4x accelerated data.

Table 2. Quantitative evaluation of ADC maps generated by DeepADC-Net and alternative state-of-the-art methods on both 4x and 8x accelerated testing datasets. Results are shown as (Mean ± Standard Deviation).

| Models | Sampling Factor | Correlation ($\times 10^{-2}$) | SSIM ($\times 10^{-2}$) | PSNR | NMSE ($\times 10^{-2}$) |
|---|---|---|---|---|---|
| Least-Squares-Fitting | 4x | 68.35 ±6.25 | 96.13 ±1.60 | 14.98 ±1.69 | 13.39 ±3.76 |
| Compressed Sensing | | 72.14 ± 6.87 | 98.79 ± 0.43 | 17.24 ± 1.44 | 7.81 ± 2.13 |
| FBPConvNet | | 87.28 ±3.66 | 99.49 ±0.15 | 21.89 ±1.11 | 2.45 ±0.34 |
| AttUnet | | 87.76 ±3.41 | 99.45 ±0.08 | 21.73 ±0.84 | 2.61 ±0.38 |
| DenseUnet | | 87.68 ±3.48 | 99.47 ±0.10 | 21.85 ±0.96 | 2.47 ±0.30 |
| Unet | | 87.41 ±3.57 | 99.48 ±0.11 | 21.89 ±1.09 | 2.46 ±0.32 |
| **DeepADC-Net** | | **90.91 ±2.28** | **99.62 ±0.06** | **23.18 ±0.90** | **1.82 ±0.19** |
| Least-Squares-Fitting | 8x | 46.00 ±7.90 | 84.86 ±4.44 | 9.60 ±1.53 | 43.15 ±9.34 |
| Compressed Sensing | | 57.56 ±22.0 | 97.36 ±3.44 | 15.90 ±2.76 | 11.46 ±10.05 |
| FBPConvNet | | 76.01 ±5.45 | 99.02 ±0.18 | 19.35 ±0.92 | 4.41 ±0.63 |
| AttUnet | | 76.76 ±5.49 | 99.00 ±0.18 | 19.30 ±0.91 | 4.52 ±0.79 |
| DenseUnet | | 76.16 ±5.31 | 99.03 ±0.19 | 19.40 ±0.92 | 4.37 ±0.64 |
| Unet | | 76.27 ±5.43 | 99.03 ±0.18 | 19.39 ±0.92 | 4.36 ±0.62 |
| **DeepADC-Net** | | **85.77 ±3.15** | **99.37 ±0.09** | **21.06 ±0.75** | **2.97 ±0.49** |

We also compared our method with a compressed sensing (CS) method (33). As summarized in Table 2, the CS method had better performance than the least-squares-fitting method but performed worse than DeepADC-Net in terms of all four performance evaluation metrics. Please refer to Supplementary Section 1.6 for more details.

Representative ADC maps of the same fully-sampled and their accelerated versions generated by all the methods under comparison are shown in Figure 3 (row 1 and row 4). The ADC maps estimated from the accelerated DW images using least-squares-fitting lose texture details when compared with their corresponding fully sampled DW images. While the CS method yielded better visualization than least-squares-fitting, the CNN based methods, including U-net, DenseU-Net, Att-UNet, and FBP-ConvNet, outperformed both CS and least-squares-fitting in generating ADC maps. Notably, DeepADC-Net generated the best accelerated ADC maps, with smaller errors than those generated by the alternative methods under comparison for both 4x and 8x accelerated image slices.

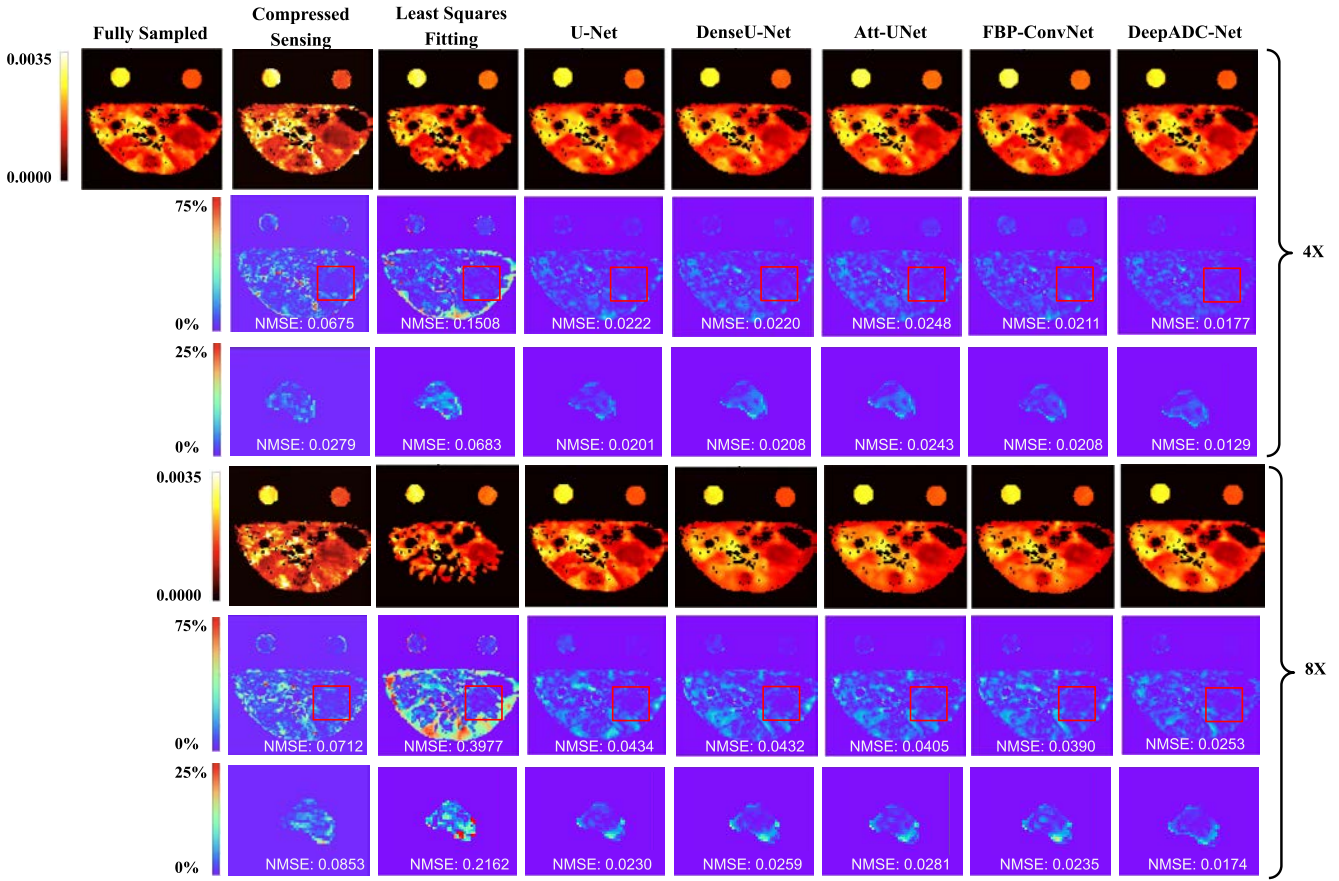

Figure 3. Visualization of 4x and 8x accelerated image slices obtained by all methods under comparison. Image slice is randomly selected with median NMSE performance. The first and fourth rows show the ground truth and the generated ADC maps. The second and fifth rows show the absolute error maps with range displayed up to 75% of the maximum difference. The third and sixth rows show the absolute error maps in the **tumor** region with range displayed up to 25% maximum difference.

### 3.2  Comparison with state-of-the-art methods on the regions of interest

We evaluated all the methods under comparison in three ROIs, including tumor, muscle, and kidney. As summarized in Table 3 for the 4x accelerated testing dataset, the least-squares-fitting showed the worst performance on all three ROIs. Additionally, the CS method had better performance than the least-squares-fitting method on all three ROIs. Among the deep learning methods under comparison, Att-UNet achieved the best performance in correlation coefficient, whereas the FBP-ConvNet yielded best performance in SSIM, PSNR and NMSE on all three ROIs. Our DeepADC-Net obtained the overall the best performance on all three ROIs. Similar trends were observed on the 8x accelerated testing dataset, as demonstrated by the results summarized in Table 4. Specifically, our DeepADC-Net obtained improved performance by 10.81%, 21.06% and 15.62% on tumor, muscle and kidney respectively in terms of correlation coefficient, compared with the second best method.

Table 3. Quantitative comparison of ADC maps generated by DeepADC-Net and alternative state-of-the-art methods for 4x accelerated testing datasets on different ROIs. Results are shown as (Mean ± Standard Deviation).

| Models | ROIs | Correlation ($\times 10^{-2}$) | SSIM ($\times 10^{-2}$) | PSNR | NMSE ($\times 10^{-2}$) |
|---|---|---|---|---|---|
| Least-Squares-Fitting | Tumor | 69.43 ±15.9 | 97.38 ±3.00 | 16.16 ±3.46 | 10.22 ±9.70 |
| Compressed Sensing | | 71.88 ±19.1 | 99.09 ±0.86 | 18.95 ±3.19 | 4.76 ±3.48 |
| FBPConvNet | | 84.62 ±9.1 | 99.50 ±0.27 | 21.44 ±2.07 | 2.50 ±1.35 |
| AttUnet | | 84.79 ±9.5 | 99.46 ±0.37 | 21.27 ±2.19 | 2.67 ±1.87 |
| DenseUnet | | 84.60 ±9.6 | 99.47 ±3.11 | 21.24 ±2.03 | 2.61 ±1.51 |
| Unet | | 84.67 ±9.1 | 99.49 ±0.28 | 21.43 ±2.10 | 2.51 ±1.26 |
| **DeepADC-Net** | | **88.37 ±6.9** | **99.62 ±0.18** | **22.49 ±1.92** | **1.90 ±0.85** |
| Least-Squares-Fitting | Muscle | 60.11 ±19.2 | 94.84 ±5.41 | 13.49 ±3.72 | 17.08 ±14.71 |
| Compressed Sensing | | 65.35 ±19.3 | 98.58 ±1.60 | 16.87 ±2.93 | 6.44 ±5.77 |
| FBPConvNet | | 79.74 ±10.0 | 99.41 ±0.31 | 20.20 ±1.83 | 2.85 ±1.26 |
| AttUnet | | 80.95 ±9.20 | 99.41 ±0.31 | 20.17 ±1.78 | 2.88 ±1.41 |
| DenseUnet | | 81.08 ±8.9 | 99.42 ±0.28 | 20.18 ±1.66 | 2.84 ±1.28 |
| Unet | | 79.94 ±9.9 | 99.41 ±0.31 | 20.17 ±1.84 | 2.87 ±1.27 |
| **DeepADC-Net** | | **86.27 ±6.0** | **99.59 ±0.19** | **21.51 ±1.50** | **2.08 ±0.84** |
| Least-Squares-Fitting | Kidney | 61.76 ±19.5 | 96.04 ±4.48 | 14.52 ±3.54 | 14.21 ±12.77 |
| Compressed Sensing | | 64.24 ±20.4 | 98.89 ±1.14 | 17.56 ±2.87 | 5.75 ±4.26 |
| FBPConvNet | | 80.40 ±12.2 | 99.45 ±2.82 | 20.67 ±1.92 | 2.74 ±1.18 |
| AttUnet | | 80.80 ±12.5 | 99.40 ±0.34 | 20.49 ±2.01 | 2.93 ±1.66 |
| DenseUnet | | 80.31 ±12.8 | 99.41 ±0.33 | 20.53 ±1.88 | 2.85 ±1.88 |
| Unet | | 80.64 ±11.7 | 99.44 ±0.28 | 20.65 ±1.89 | 2.73 ±1.12 |
| **DeepADC-Net** | | **85.30 ±9.7** | **99.57 ±0.20** | **21.76 ±1.76** | **2.11 ±0.87** |

Figure 3 (row 3 and row 6) also shows ADC error maps in a tumor region for both 4x and 8x accelerated data, showing that the direct least-squares-fitting of accelerated images yielded the worst performance. All the deep learning methods obtained improved performance, and our method obtained the overall best performance.

Table 4. Quantitative comparison of ADC maps on DeepADC-Net and alternative state-of-the-art methods for 8x accelerated testing datasets on different ROIs. Results are shown as (Mean ± Standard Deviation).

| Models | ROIs | Correlation ($\times 10^{-2}$) | SSIM ($\times 10^{-2}$) | PSNR | NMSE ($\times 10^{-2}$) |
|---|---|---|---|---|---|
| Least-Squares-Fitting | | 45.64 ±18.5 | 89.24 ±7.92 | 10.51 ±3.11 | 33.84 ±20.72 |
| Compressed Sensing | | 58.18 ±22.3 | 97.25 ±5.27 | 16.14 ±3.89 | 10.79 ±13.10 |
| FBPConvNet | | 72.11 ±14.0 | 99.10 ±0.51 | 18.99 ±2.52 | 4.42 ±2.52 |
| AttUnet | Tumor | 72.68 ±14.2 | 99.06 ±0.53 | 18.85 ±2.01 | 4.63 ±2.89 |
| DenseUnet | | 73.01 ±16.3 | 99.16 ±0.46 | 19.17 ±1.97 | 4.17 ±2.16 |
| Unet | | 71.94 ±14.0 | 99.09 ±0.50 | 19.00 ±1.97 | 4.42 ±2.52 |
| **DeepADC-Net** | | **83.82 ±9.7** | **99.45 ±0.31** | **21.08 ±2.00** | **2.76 ±1.32** |
| Least-Squares-Fitting | | 27.32 ±17.10 | 75.14 ±9.48 | 6.24 ±1.84 | 68.55 ±19.74 |
| Compressed Sensing | | 49.77 ±20.9 | 95.81 ±6.84 | 13.85 ±3.36 | 14.23 ±16.53 |
| FBPConvNet | | 59.19 ±17.5 | 98.92 ±0.48 | 17.57 ±1.62 | 5.11 ±1.93 |
| AttUnet | Muscle | 59.64 ±17.2 | 98.88 ±0.55 | 17.48 ±1.68 | 5.27 ±2.43 |
| DenseUnet | | 58.80 ±17.5 | 98.91 ±0.21 | 17.55 ±1.60 | 5.13 ±1.97 |
| Unet | | 59.23 ±16.7 | 98.92 ±0.46 | 17.55 ±1.58 | 5.12 ±1.91 |
| **DeepADC-Net** | | **80.70 ±7.4** | **99.41 ±0.30** | **19.63 ±1.55** | **2.94 ±1.76** |
| Least-Squares-Fitting | | 34.10 ±16.8 | 80.11 ±10.11 | 7.58 ±2.28 | 56.61 ±22.48 |
| Compressed Sensing | | 49.99 ±22.5 | 96.53 ±5.48 | 14.55 ±3.57 | 13.34 ±14.81 |
| FBPConvNet | | 62.29 ±17.7 | 98.91 ±0.60 | 17.99 ±1.92 | 5.16 ±2.61 |
| AttUnet | Kidney | 62.36 ±17.5 | 98.85 ±0.66 | 17.84 ±1.92 | 5.42 ±3.23 |
| DenseUnet | | 61.99 ±17.8 | 98.91 ±5.63 | 18.01 ±1.92 | 5.13 ±2.47 |
| Unet | | 62.53 ±17.1 | 98.91 ±0.57 | 18.01 ±1.94 | 5.12 ±2.43 |
| **DeepADC-Net** | | **78.15 ±11.5** | **99.30 ±0.32** | **19.73 ±1.80** | **3.43 ±1.69** |

### 3.3 Ablation studies on network inputs, loss functions and self-attentions

Table 5. Ablation studies of ADC maps for 4x accelerated testing datasets. Results are shown as (Mean ± Standard Deviation).

| Models | Correlation ($\times 10^{-2}$) | SSIM ($\times 10^{-2}$) | PSNR | NMSE ($\times 10^{-2}$) |
|---|---|---|---|---|
| DenseU-Net | 87.68 ± 3.48 | 99.47 ± 0.10 | 21.85 ± 0.96 | 2.47 ± 0.30 |
| DenseU-ADC | 90.15 ± 2.65 | 99.59 ± 0.06 | 22.81 ± 0.91 | 1.96 ± 0.21 |
| DenseU-DWI | 84.37 ± 3.13 | 99.37 ± 0.10 | 20.80 ± 0.70 | 3.15 ± 0.48 |
| DenseU-ADC-DWI | 90.63 ± 2.46 | 99.62 ± 0.07 | 23.05 ± 0.88 | 1.87 ± 0.24 |
| **DeepADC-Net** | **90.91 ± 2.28** | **99.62 ± 0.06** | **23.18 ± 0.90** | **1.82 ± 0.19** |

Table 5 summarizes quantitative performance measures of the deep learning models built by our method with different components. Specifically, DenseU-Net had the overall worst performance and its performance was worse than DenseU-ADC, indicating that the multi-channel input of both the accelerated DW images and their ADC map provided richer information than the accelerated ADC map alone for generating high-quality ADC maps.

The deep learning models with the same multi-channel input but built by the proposed method with different loss function terms had a different performance. Specifically, DenseU-DWI had the worst performance compared with all the other two models that were trained by optimizing three complementary loss function terms, including $L_{adc}$, $L_{s_0}$, and $L_{dwi}$, whereas DenseU-DWI was trained by optimizing $L_{dwi}$ alone. DenseU-ADC-DWI shared with DeepADC-Net the same input and the same loss function, but it performed worse than DeepADC-Net, indicating that the multi-head self-attention module was useful to improve the ADC generation.

## 4   Discussion

This study has demonstrated that a new deep leaning method, referred to as DeepADC-Net, successfully generated ADC maps from accelerated diffusion-weighted MR data. The network contained three complementary loss functions, including the difference between ADC maps learned and those computed from fully-sampled DW images ($L_{adc}$) and the differences between fully-sampled DW images and derived DW images from the learned $S_0$ and ADC maps ($L_{S_0}$ and $L_{dwi}$). The network was further enhanced by a bottleneck transformer with multi-head self-attention module. The method has been evaluated on accelerated DW images by 4x and 8x. Quantitative performance measures have demonstrated that our method obtained an accurate estimation of ADC maps, which reduces the acquisition time from 25 minutes to just over six minutes. Since DW-MRI is highly sensitive to any unwanted motion that is not diffusion-related and motion is often exacerbated by long scan times, such reduction in scan time could potentially yield more accurate ADC values and lead to better distinction between benign and malignant tumors or to better assessment of changes in response to treatment.

We have compared our method with conventional least-squares-fitting and state-of-the-art deep learning methods. In particular, we compared our method with the least-squares-fitting algorithm, U-Net (30), DenseU-Net (31), FBP-ConvNet (16), and Att-UNet (17) with the same training and testing settings. Comparison results, as summarized in Tables 2, 3, and 4, have demonstrated that DeepADC-Net obtained the best results among all methods under comparison for ADC map generation on both the entire cross-section images and specific ROIs. Figure 3 visualizes differences between ADC maps generated by DeepADC-Net and the alternative methods under comparison, demonstrating that DeepADC-Net generated high-quality ADC maps with the minimum errors and less discrepancy from full-sampled ADC maps. Furthermore, DeepADC-Net outperformed the CS method in generating ADC maps as demonstrated in Table 2, while the CS method performed better than the least-squares-fitting method. This shows that DL can overcome the limitations of the CS method, such as inefficiency in capturing complex features through sparsifying transforms (34,35).

The present study built deep learning models to compute high quality ADC maps from accelerated DW images, different from typical image super-resolution studies. Though many deep learning algorithms have been developed for accelerated MRI data, our method was applied to a dataset obtained by radially sampled diffusion weighted spin-echo (Rad-DW-SE) acquisition method for quantitative ADC estimation. Our approach is fundamentally different from existing DL methods that focus on improving MR images from accelerated k-space data. Our method directly computes ADC maps in the deep learning process with regularization terms built upon the monoexponential model, facilitating end-to-end learning of the ADC maps with improved quality and efficiency. It is noteworthy that our proposed model can be considered as a plug-and-play module that can be adopted by other deep learning approaches, including those under comparison. Since no alternative deep learning method is available for a direct comparison, we evaluated the proposed method through extensive ablation studies. Our ablation studies have demonstrated that 1) the accelerated DWI and their derived ADC maps as multi-channel inputs could improve the model performance compared to the model with only ADC maps as its input, 2) minimizing the discrepancy between the DW images computed from the generated ADC maps and fully-sampled DW images could regularize the DL-generated ADC maps, 3) the bottleneck transformer was useful to improve the generation of ADC maps. Additionally, we have evaluated DeepADC-Net on both real-world and simulated 4x accelerated datasets (Supplementary Section 1.1, Table S1) and experimental results show that DeepADC-Net outperformed standard curve-fitting methods on real-world accelerated data. However, the results should be interpreted with a caveat that the real-world 4x accelerated data might capture diffusion information different from that captured by the full-sampled data in that they were not collected simultaneously.

Our research has several limitations. First, our ablation studies were carried out based on the 4x accelerated dataset that we tune model parameters by fixing some of them while testing the rest, instead of using a fine-grid searching method. Second, we trained and evaluated our method largely on simulated data at two different acceleration factors since it is difficult to collect both real-world accelerated and fully-sampled data simultaneously. We did evaluate our method based on real-world accelerated data and compared their derived ADC maps with those computed from fully-sampled data subsequently. However, such pairs of data might capture different diffusion information. Third, the present study only considered identical, evenly spaced view angles of DW images at all b-values. A more complete analysis would involve other view ordering schemes, such as golden-angle, or one in which different view angles are used to encode for different b-values. In our current work, although fewer b-values are required to fit an exponential curve, five b-values were collected in order to improve least-square fitting and also to leave open the possibility to access diffusion kurtosis in subsequent analysis. The availability of additional b-values may have facilitated high acceleration factors that were achieved in this work than in fewer b-values had been available. Lastly, our research has been focusing on preclinical imaging studies. Therefore, we have not specifically tested our method on human data, such as those collected using PROPELLER (36). We believe that our method can be applied to any other data with different image acquisition schemes given that the proposed method is based on deep learning, which can learn the underlying relationships among k-space data, regardless of the acquisition scheme used to obtain the data. The network, however, would need to be retrained with data acquired using the desired acquisition scheme.

# 5  Conclusion

We developed a deep learning method, referred to as DeepADC-Net, to generate apparent diffusion coefficient maps from accelerated diffusion-weighted MR data, achieving 4 to 8-fold acceleration of DWI acquisition. The proposed DeepADC-Net integrating a densely connected Encoder-Decoder architecture with a vision transformer is shown to perform superior to widely used compressed sensing and several state-of-the-art deep learning models for computing ADC maps.

# Acknowledgements

This work was supported in part by the National Institutes of Health [grant numbers: MH120811, EB022573, AG066650, and U24 CA231858 (Penn Pancreatic Cancer Imaging Resource)].

# Learning Apparent Diffusion Coefficient Maps from Accelerated Radial k-Space Diffusion-Weighted MRI in Mice using a Deep CNN-Transformer Model

Yuemeng Li[1,2], Miguel Romanello Joaquim[2], Stephen Pickup[2], Hee Kwon Song[2], Rong Zhou[2,3], Yong Fan[1,2]*

[1]Center for Biomedical Image Computing and Analytics (CBICA), Perelman School of Medicine, University of Pennsylvania, Philadelphia, PA 19104, USA
[2]Department of Radiology, Perelman School of Medicine, University of Pennsylvania, Philadelphia, PA 19104, USA
[3]Abramson Cancer Center, University of Pennsylvania, Philadelphia, PA 19104, USA

## Supplementary data

We have made our code publicly available at GitHub: https://github.com/ymli39/DeepADC-Net-Learning-Apparent-Diffusion-Coefficient-Maps, and our dataset can be downloaded at https://pennpancreaticcancerimagingresource.github.io/data.html.

### 1.1. Evaluation on real-world testing data

To investigate the generalizability of our proposed DeepADC-Net on real-world dataset, we acquired **nine additional scans** using one quarter of the original views as *real-world* 4x accelerated data (101 views). The fully sampled data (403 views) were sequentially collected with the same diffusion weighted, radially sampled spin echo sequence (Rad-DW-SE) with even sampling of the view angle over 360 degrees (FOV= $32 \times 32$ mm$^2$, $slices = 8 - 19, 96$ readout points, slice thickness= $1.5\ mm$, TR= $750\ msec$, TE=$28.7\ msec$, $b$-values= $10, 535, 1070, 1479, 2141\ s/mm^2$). The **simulated** 4x accelerated data were generated by sampling one out of every four radial views in k-space from the fully-sampled data, which is the same procedure used in other experiments of this study. It is worth noting that ***the real-world 4x accelerated data might capture diffusion information different from that captured by the full-sampled data*** in that they were not collected simultaneously.

DeepADC-Net was trained on 2668 slices of 174 animals with both fully sampled and simulated accelerated scans, and tested on 144 slices of both real-world and simulated 4x accelerated data collected from 9 animals. We applied DeepADC-Net model to both the real-world and simulated 4x accelerated data to generate ADC maps, which were compared to those computed from the fully-sampled data using least-square fitting of the monoexponential model in Equation (1). Additionally, we applied compressed sensing to both the real-world and simulated 4x accelerated data to generate reconstructed ADC maps, allowing for a comprehensive comparison.

Table S1. Quantitative comparison of ADC maps on DeepADC-Net for both simulated and real-world 4x accelerated testing datasets. Results are shown as (Mean ± Standard Deviation).

| Models | Dataset | Correlation ($\times 10^{-2}$) | SSIM ($\times 10^{-2}$) | PSNR | NMSE ($\times 10^{-2}$) |
|---|---|---|---|---|---|
| Least-Squares-Fitting | Real-world | 60.26 ± 16.30 | 93.23 ± 8.95 | 13.67 ± 3.21 | 21.83 ± 22.56 |
|  | Simulated | 69.49 ± 14.88 | 96.33 ± 2.79 | 15.11 ± 2.43 | 14.06 ± 13.46 |
| Compressed Sensing | Real-world | 59.51 ± 20.49 | 97.87 ± 1.81 | 16.22 ± 2.31 | 11.19 ± 12.95 |
|  | Simulated | 66.15 ± 18.39 | 98.09 ± 1.69 | 16.98 ± 2.76 | 9.33 ± 8.76 |
| DeepADC-Net | Real-world | 78.06 ± 14.73 | 98.71 ± 1.87 | 19.53 ± 2.80 | 5.75 ± 8.21 |
|  | Simulated | 89.17 ± 7.97 | 99.42 ± 1.09 | 22.46 ± 2.15 | 3.00 ± 5.61 |

As summarized in Table S1, the ADC maps generated by the deep learning model from both the real-world and simulated data were similar to those computed from the fully-sampled data. On the simulated accelerated dataset, the ADC maps computed from the fully-sampled data were closer to those generated by the deep learning model than those computed using the least-square fitting and compressed sensing methods, indicating that the deep learning model can generate ADC maps better than the standard ADC computing methods. The compressed sensing method yielded worse performance in terms of correlation coefficient but better performance in terms of SSIM, PSNR and NMSE when compared to least-square fitting method. A similar trend was observed on the real-world accelerated dataset, and the ADC maps generated by the deep learning model were closer to those computed from the fully-sample data, though they might capture different diffusion information.

### 1.2. Tuning of regularization parameters

The regularization parameters ($\beta$ and $\gamma$) of our loss function were tuned with $\alpha$=1 on the 4x accelerated dataset. We first built a model without the regularization functions of $L_{S_0}$ and $L_{DWI}$, i.e., $\beta = \gamma = 0$, and then built models with varied $\beta$ and $\gamma$, selected from 0.1, 0.5 and 1. The testing results summarized in Table S2 indicated that $\beta = \gamma = 0.1$ yielded the overall best performance in terms of the correlation coefficient, PSNR and NMSE, whereas $\beta = \gamma = 0.5$ yielded better results in terms of SSIM. Therefore, we utilized weight $\beta = \gamma = 0.1$ throughout the experiments.

Table S2. Performance of DeepADC-Net trained with different regularization parameters on the 4x accelerated testing dataset. Results are shown as (Mean ± Standard Deviation)

| Weights | Correlation ($\times 10^{-2}$) | SSIM ($\times 10^{-2}$) | PSNR | NMSE ($\times 10^{-2}$) |
|---|---|---|---|---|
| $\alpha = 1, \beta = 0, \gamma = 0$ | 87.68 ± 3.48 | 99.47 ± 0.10 | 21.85 ± 0.96 | 2.47 ± 0.30 |
| $\alpha = 1, \beta = 0.1, \gamma = 0.1$ | **90.91 ± 2.28** | 99.62 ± 0.06 | **23.18 ± 0.90** | **1.82 ± 0.19** |
| $\alpha = 1, \beta = 0.5, \gamma = 0.5$ | 90.86 ± 2.57 | **99.63 ± 0.07** | 23.17 ± 1.01 | 1.82 ± 0.25 |
| $\alpha = 1, \beta = 1, \gamma = 1$ | 90.79 ± 2.63 | 99.62 ± 0.08 | 23.09 ± 1.05 | 1.86 ± 0.29 |

### 1.3. Ablation studies on tumor, muscle, and kidney

The same ablation studies were carried out to evaluate how the deep learning models built by our method with different components performed on ROIs, including tumor, muscle, and kidney, based on both 4x and 8x accelerated datasets. As summarized in Tables S3 and S4, the input of both ADC map and DW images led to better performance, similar to those results obtained on the whole images. Similarly, deep learning models (DenseU-ADC-DWI and DeepADC-Net) trained by optimizing the combination loss function of $L_{adc}$, $L_{S_0}$ and $L_{dwi}$ had better performance than those trained by optimizing $L_{adc}$ or $L_{dwi}$ alone. The self-attention module did consistently improve the performance in terms of correlation coefficient, but yielded degraded performance in terms of SSIM, PSNR, or NMSE, particularly on the data with 8x acceleration factor. We postulated that such degradation might be attributed to the fact that all the model parameters were tuned on the 4x accelerated data based on the performance estimated on the whole image basis.

Table S3. Ablation studies of ADC maps of the 4x accelerated testing dataset on different ROIs. Results are shown as (Mean ± Standard Deviation).

| Models | ROIs | Correlation ($\times 10^{-2}$) | SSIM ($\times 10^{-2}$) | PSNR | NMSE ($\times 10^{-2}$) |
|---|---|---|---|---|---|
| DenseU-Net | Tumor | 84.60 ±9.6 | 99.47 ±3.11 | 21.24 ±2.03 | 2.61 ±1.51 |
| DenseU-ADC | | 87.79 ±7.3 | 99.60 ±0.18 | 22.29 ±1.86 | 1.98 ±0.82 |
| DenseU-DWI | | 81.72 ±9.4 | 99.43 ±0.30 | 20.53 ±1.83 | 3.01 ±1.44 |
| DenseU-ADC-DWI | | 88.18 ±6.9 | 99.62 ±0.19 | **22.49 ±1.89** | 1.91 ±0.85 |
| DeepADC-Net | | **88.37 ±6.9** | **99.62 ±0.18** | 22.49 ±1.92 | **1.90 ±0.85** |

| Models | | | | | |
|---|---|---|---|---|---|
| DenseU-Net | | 81.08 ±8.9 | 99.42 ±0.28 | 20.18 ±1.66 | 2.84 ±1.28 |
| DenseU-ADC | | 85.36 ±6.5 | 99.55 ±0.20 | 21.21 ±1.50 | 2.21 ±0.86 |
| DenseU-DWI | Muscle | 77.35 ±1.3 | 99.37 ±0.33 | 19.46 ±1.48 | 3.30 ±1.27 |
| DenseU-ADC-DWI | | 86.18 ±5.8 | 99.58 ±0.20 | 21.37 ±1.50 | 2.11 ±0.90 |
| DeepADC-Net | | **86.27 ±6.0** | **99.59 ±0.19** | **21.51 ±1.50** | **2.08 ±0.84** |
| DenseU-Net | | 80.31 ±12.8 | 99.41 ±0.33 | 20.53 ±1.88 | 2.85 ±1.88 |
| DenseU-ADC | | 84.23 ±10.7 | 99.55 ±0.08 | 21.48 ±1.71 | 2.21 ±0.85 |
| DenseU-DWI | Kidney | 77.00 ±12.2 | 99.37 ±0.31 | 19.79 ±1.67 | 3.27 ±1.67 |
| DenseU-ADC-DWI | | 84.99 ±10.2 | **99.58 ±0.19** | 21.73 ±1.68 | **2.10 ±0.93** |
| DeepADC-Net | | **85.30 ±9.7** | 99.57 ±0.20 | **21.76 ±1.76** | 2.11 ±0.87 |

Table S4. Ablation studies of ADC maps of the 8x accelerated testing dataset on different ROIs. Results are shown as (Mean ± Standard Deviation).

| Models | ROIs | Correlation ($\times 10^{-2}$) | SSIM ($\times 10^{-2}$) | PSNR | NMSE ($\times 10^{-2}$) |
|---|---|---|---|---|---|
| DenseU-Net | | 73.01 ±16.3 | 99.16 ±0.46 | 19.17 ±1.97 | 4.17 ±2.16 |
| DenseU-ADC | | 81.18 ±10.1 | 99.37 ±0.31 | 20.47 ±1.83 | 3.06 ±1.41 |
| DenseU-DWI | Tumor | 72.93 ±12.7 | 99.15 ±0.41 | 18.66 ±1.85 | 4.65 ±2.75 |
| DenseU-ADC-DWI | | 82.04 ±10.5 | 99.43 ±0.26 | 20.70 ±1.86 | 2.85 ±1.19 |
| DeepADC-Net | | **83.82 ±9.7** | **99.45 ±0.31** | **21.08 ±2.00** | **2.76 ±1.32** |
| DenseU-Net | | 58.80 ±17.5 | 98.91 ±2.14 | 17.55 ±1.60 | 5.13 ±1.97 |
| DenseU-ADC | | 78.52 ±8.2 | 99.34 ±0.29 | 19.57 ±1.46 | 3.24 ±1.46 |
| DenseU-DWI | Muscle | 62.60 ±10.0 | 98.94 ±0.44 | 17.16 ±1.25 | 5.40 ±1.51 |
| DenseU-ADC-DWI | | 79.71 ±7.7 | **99.41 ±0.27** | **19.82 ±1.44** | 2.98 ±1.15 |
| DeepADC-Net | | **80.70 ±7.4** | 99.41 ±0.30 | 19.63 ±1.55 | **2.94 ±1.76** |
| DenseU-Net | | 61.99 ±17.8 | 98.91 ±5.63 | 18.01 ±1.92 | 5.13 ±2.47 |
| DenseU-ADC | | 77.13 ±11.9 | 99.30 ±0.33 | 19.74 ±1.68 | 3.42 ±1.81 |
| DenseU-DWI | Kidney | 62.96 ±14.6 | 99.03 ±0.44 | 17.69 ±1.54 | 5.11 ±2.03 |
| DenseU-ADC-DWI | | 77.94 ±11.6 | **99.33 ±0.33** | **19.91 ±1.75** | **3.24 ±1.44** |
| DeepADC-Net | | **78.15 ±11.5** | 99.30 ±0.32 | 19.73 ±1.80 | 3.43 ±1.69 |

**1.4. Visualization of ADC maps on muscle and kidney**

Figures S1 (muscle) and S2 (kidney) show ADC maps generated by different methods under comparison to highlight muscle and kidney, respectively. Similar to those maps shown in Figure S1, all the deep learning methods had better performance than the standard ADC computing method, i.e., the least-square-fitting, and DeepADC-Net generated ADC maps with the smallest error and highest similarity to the ground truth.

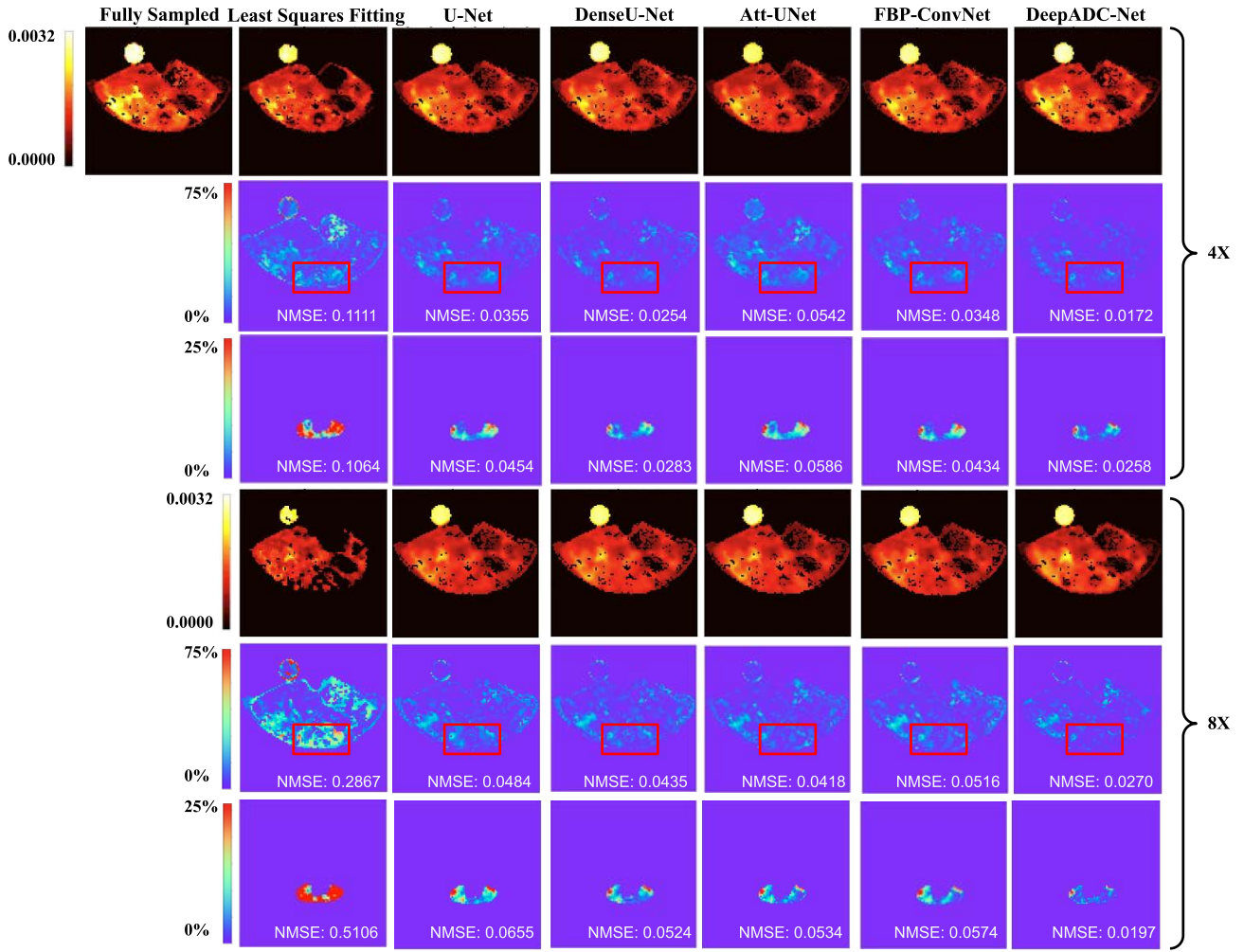

Figure S1. Representative cases of 4x and 8x accelerated ADC maps obtained by the methods under comparison. The first and fourth rows show the ground truth (fully-sampled) and the generated ADC maps. The second and fifth rows show the absolute error maps with range displayed up to 75% of the maximum difference. The third and sixth rows show the absolute error maps in the **muscle** region with range displayed up to 25% maximum difference.

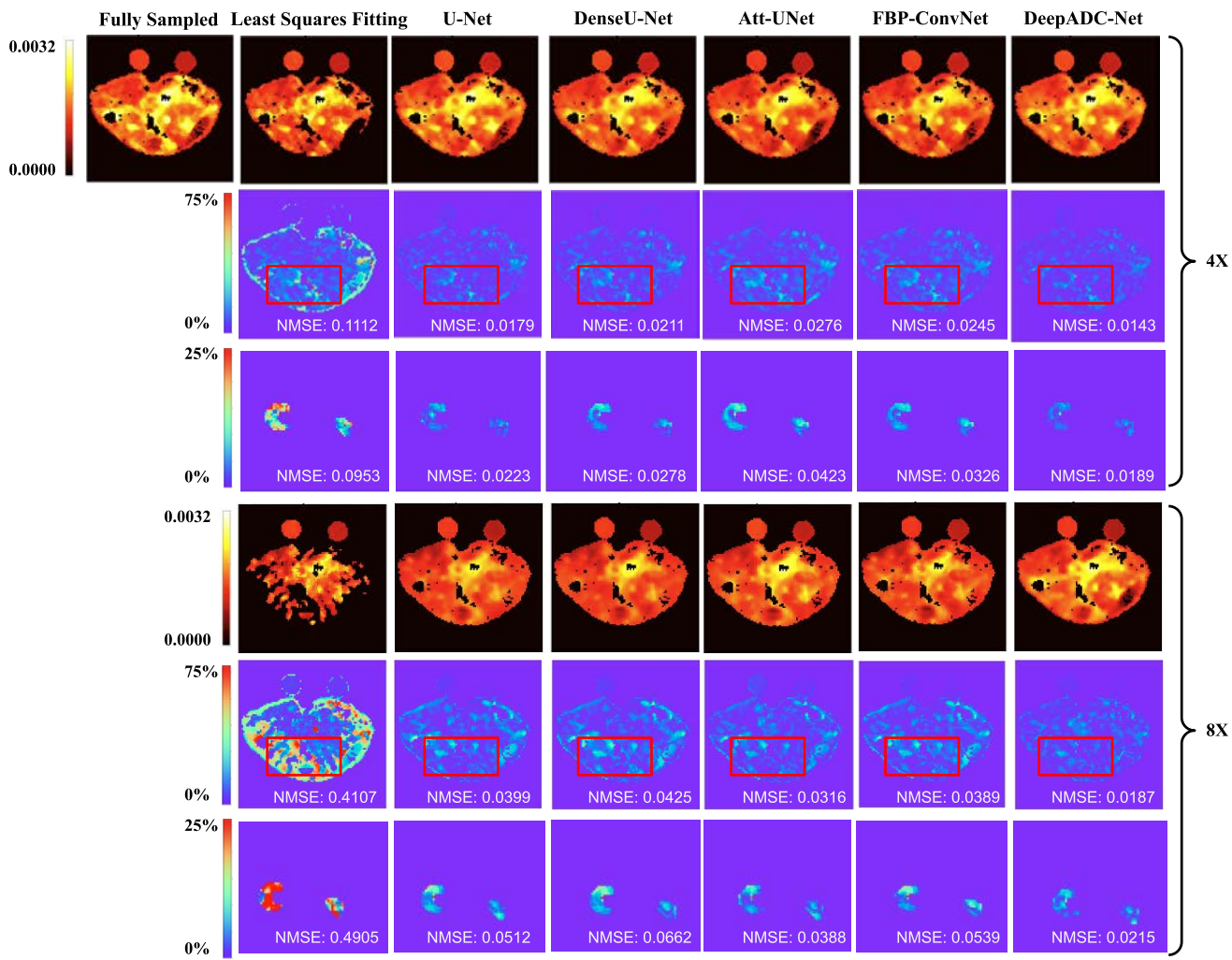

Figure S2. Representative cases of 4x and 8x accelerated ADC maps obtained by the methods under comparison. The first and fourth rows show the ground truth (fully-sampled) and the generated ADC maps. The second and fifth rows show the absolute error maps with range displayed up to 75% of the maximum difference. The third and sixth rows show the absolute error maps in the **kidney** region with range displayed up to 25% maximum difference.

### 1.5. Ablation studies on numbers of filter sizes and convolutional layers

We evaluated the impact of the number of convolutional layers and kernel sizes on the performance of our deep learning model. We varied the filter size of each densely connected block in all encoders and decoders with sizes of 64, 128, 196, and 256. We then fixed the filter size to test the effectiveness of different numbers of consecutive convolutional layers in each densely connected block. In these ablation studies summarized in Table S5, all models were trained with the same training setting with the presence of 1) the DW images and ADC maps as multi-channel inputs, 2) supervisions on ADC maps, $S_0$ and DW images ($L_{adc}$, $L_{S_0}$ and $L_{dwi}$) during training and 3) the transformer module at network bottleneck.

As summarized in the top four rows of Table S5, filter size of 128 yielded the best performance for correlation coefficient; 256 performed best for SSIM; and 64 was best for PSNR and NMSE. We adopted a filter size of 64 for our proposed network due to its computational efficiency and overall competitive performance. The bottom two rows of Table S5 summarize the performance of DeepADC-Net models trained with different numbers of consecutive convolutional layers inside each densely connected block. We evaluated the network backbone with three and five consecutive convolutional layers, and these results indicated that the model with three consecutive convolutional layers yielded the best results for SSIM, PSNR and NMSE. We therefore adopted a convolution layers size of 3 for our proposed network.

Table S5. Ablation studies of ADC maps with different filter sizes and with different numbers of convolutional layers for 4x accelerated testing datasets. Results are shown as (Mean ± Standard Deviation).

| Convolution Layers | Filter Size | Correlation($\times 10^{-2}$) | SSIM ($\times 10^{-2}$) | PSNR | NMSE ($\times 10^{-2}$) |
|---|---|---|---|---|---|
| 3 | 64 | 90.91 ±2.28 | 99.62 ±0.06 | **23.18 ±0.90** | **1.82 ±0.19** |
| 3 | 128 | **90.93 ±2.53** | 99.62 ±0.07 | 23.08 ±1.02 | 1.85 ±0.25 |
| 3 | 196 | 90.63 ±2.64 | 99.60 ±0.07 | 22.96 ±0.89 | 1.93 ±0.37 |
| 3 | 256 | 90.69 ±2.67 | **99.63 ±0.08** | 23.10 ±1.07 | 1.86 ±0.29 |
| 3 | 64 | 90.91 ±2.28 | **99.62 ±0.06** | **23.18 ±0.90** | **1.82 ±0.19** |
| 5 | 64 | **90.97 ±2.58** | 99.61 ±0.08 | 23.04 ±1.06 | 1.87 ±0.27 |

**1.6. Compressed sensing for 8x accelerated dataset**

A simple out-of-the-box CS reconstruction method using the SparseMRI library improved the ADC maps for 4x accelerated data, as shown in Table 2 and Table 3. The same method did not reliably improve every scan in the 8x accelerated dataset. Therefore, for the 8x accelerated CS method, we optimized the number of CS iterations for each b-value using a random sample of 10 scans from the testing dataset and comparing their mean MSE relative to fully-sampled DWIs after each iteration, shown in Figure S3. Eight scans reached a steady-state in 50 iterations, all of which were at their lowest error or slightly worse. Of the two remaining scans, one was decreasing its error with subsequent CS iterations, while the other was increasing. This trend was observed for all five b-values. 50 iterations was chosen as an optimum method for reconstruction of the 8x accelerated dataset. For one of the 36 testing scans, the optimized CS method for 8x accelerated data still produced a significantly worse image than was inputted, so the scan was removed from the results.

As shown in Table 2 for whole slice evaluation, the 8x accelerated datasets demonstrated that the CS method outperformed the least-square-fitting method across all four criteria, including Correlation coefficient, SSIM, PSNR and NMSE, while all DL methods yielded better performance compared to CS method. Furthermore, Table 4 illustrates the reconstructing of specific ROIs on the 8x accelerated dataset, where CS method showed a significant performance improvement over the least-square-fitting method. The DL methods outperformed the CS method across all ROIs.

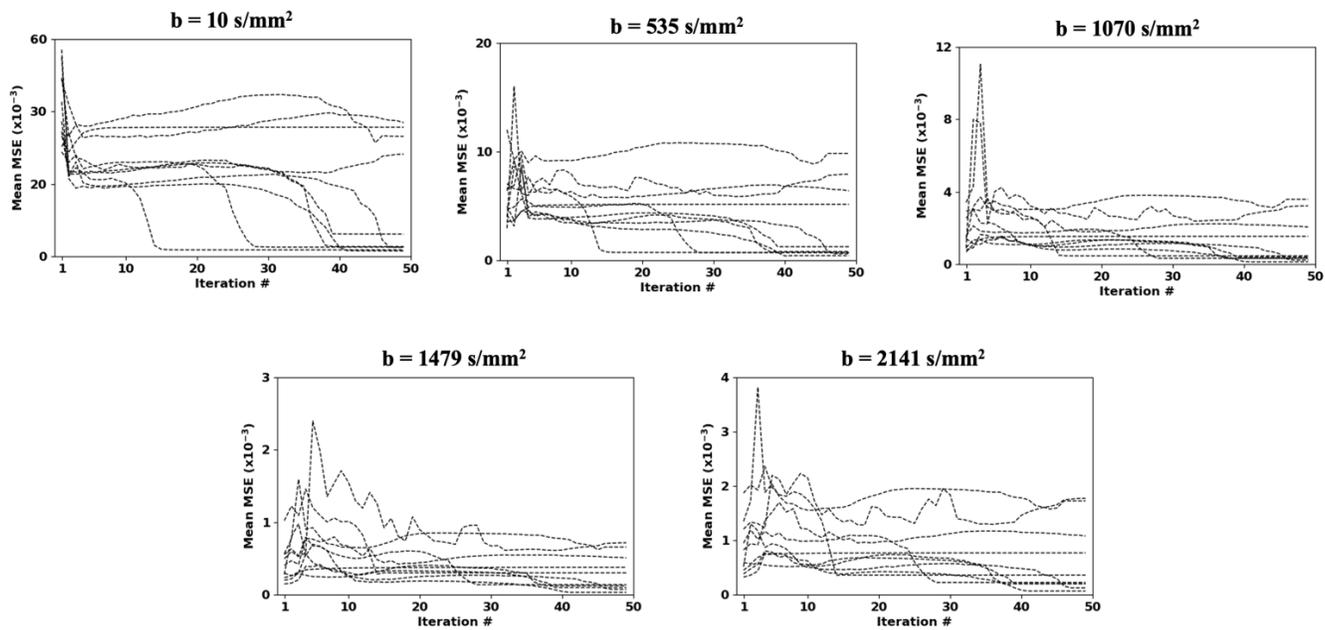

Figure S3. Mean MSE of a random set of 10 8x accelerated scans as a function of number of CS iterations for each b-value. Fully-sampled DWIs were used as ground truth.